\documentclass[11pt,a4paper]{article}
\usepackage[latin1]{inputenc}
\usepackage[english]{babel}
\usepackage{graphicx}
\usepackage{mathtext}
\usepackage{pstricks}
\usepackage{color}
\usepackage{epsfig,cite}
\usepackage{amsmath}
\usepackage{amssymb}
\usepackage{amsfonts}
\usepackage{latexsym}
\usepackage{wasysym}
\usepackage{bm}
\catcode`@=11 

\newcommand{\beq}{\begin{equation}}
\newcommand{\eeq}{\end{equation}}
\newcommand{\bea}{\begin{eqnarray}}
\newcommand{\eea}{\end{eqnarray}}
\newcommand{\bdm}{\begin{displaymath}}
\newcommand{\edm}{\end{displaymath}}
\def\as{\alpha_s}

\def \msb{\overline{\textrm{MS}}}
\def\d{\partial}

\def \d{{\rm d} }

\newcommand{\lp}{\left(}
\newcommand{\rp}{\right)}

\begin{document}
\pagestyle{empty}

\begin{flushright}
MAN/HEP/2011/08 \\
\end{flushright}

\begin{center}
\vspace*{2.5cm}
{\Large \bf High energy resummation for rapidity distributions$^*$}
 \\
\vspace{0.3cm}
Simone Marzani$^a$, Fabrizio Caola$^{b,c}$, and Stefano Forte$^b$
\\
\vspace{0.3cm}  {\it
{}$^a$School of Physics \& Astronomy, University of Manchester,\\
Oxford Road, Manchester, M13 9PL, England, U.K.\\ \medskip
{}$^b$Dipartimento di Fisica, Universit\`a di Milano and \\
INFN, Sezione di Milano, I-20133 Milano, Italy\\ \medskip
{}$^c$Fermilab, Batavia, IL 60510, USA}
\vspace*{1.5cm}

{\bf Abstract}
\end{center}

\noindent

We discuss the generalisation of  high-energy resummation to rapidity distributions to leading logarithmic accuracy. 
 We test our procedure applying it to Higgs production in gluon-gluon fusion both with finite top mass and in the infinite mass limit. We check that they reproduce the known results at fixed order and we estimate finite top mass corrections to the NLO distribution.
\vspace*{1cm}

\vfill
\noindent

\begin{flushleft} $^*$Presented at the XIX International Workshop on Deep-Inelastic Scattering and Related Subjects (DIS 2011), April 11-15, 2011; Newport News, Virginia, USA
\end{flushleft}
\eject

\setcounter{page}{1} \pagestyle{plain}

\section{Introduction}

The resummation of leading high energy (or small-$x$) contributions to hard QCD processes has been known for a long time in the case of inclusive cross-sections:
 heavy quark photo- and
lepto-production~\cite{CataniHQ}, Deep-Inelastic
Scattering~\cite{CataniDIS}, and more recently hadroproduction
processes, including heavy quarks~\cite{BallEllis},
Standard Model~\cite{HautmannHiggs,SimoneHiggs,SimoneHiggsProc} and pseudo-scalar~\cite{pseudo} Higgs production in gluon-gluon fusion,
Drell-Yan~\cite{SimoneDY} and prompt-photon~\cite{GiovanniPhoton}.
In Ref.~\cite{CFM} we generalised the resummation formalism to rapidity distributions, opening the possibility to studies which are more relevant from a phenomenological viewpoint.

\begin{figure}
\begin{center}
\includegraphics[height=.3\textheight]{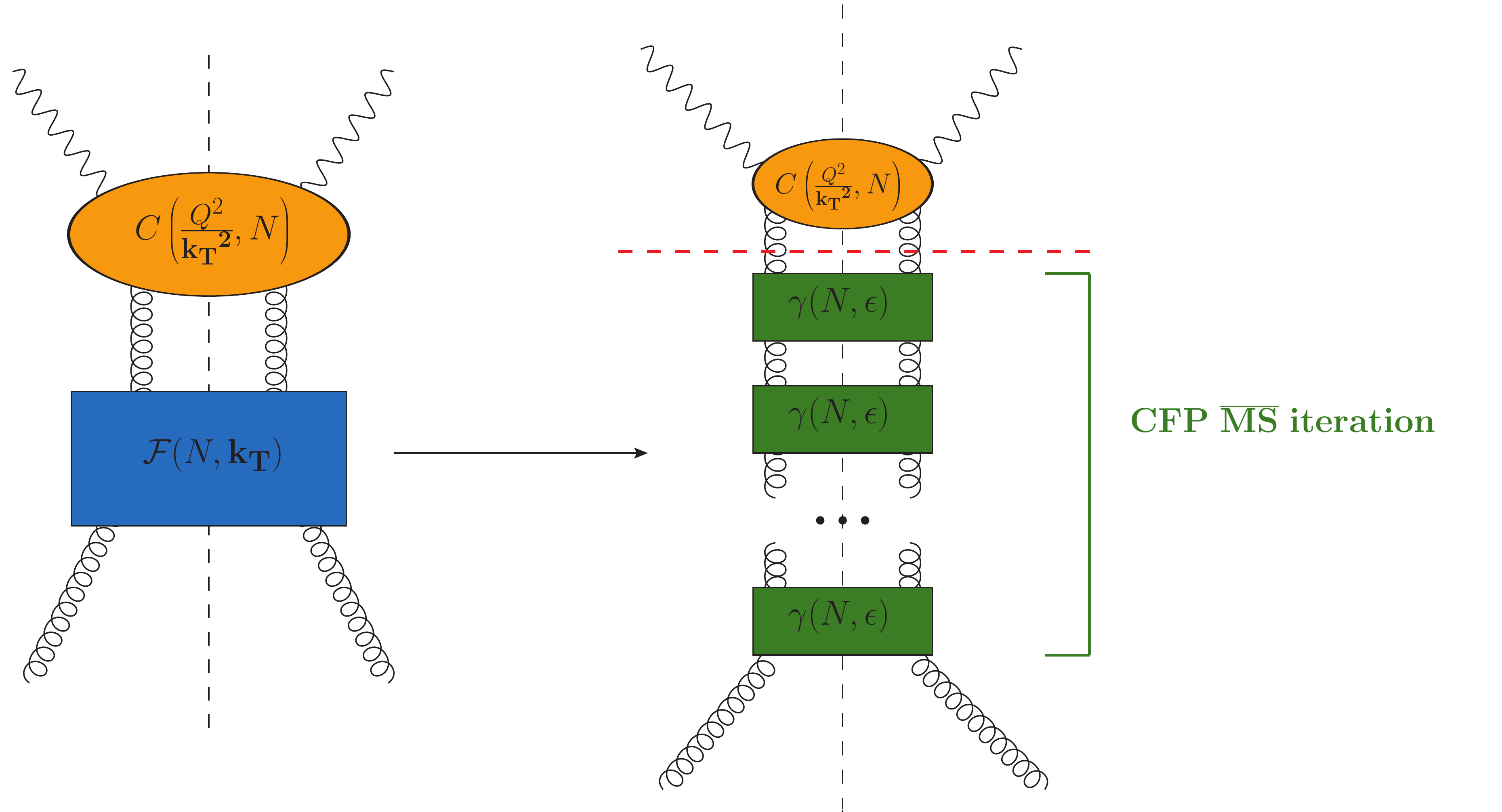}
\caption{Left: decomposition of the partonic cross-section in terms of
  two-gluon irreducible hard part and a reducible ladder part. Right:
generalized ladder expansion of the ladder part.}\label{fig:cfp}
\end{center}
\end{figure}

Such a result was made possible by a different, but equivalent, approach to small-$x$ resummation. In the standard approach~\cite{CataniHQ} one writes the cross-section in the high-energy limit as a convolution of a two-gluon irreducible hard part  and a reducible ladder part, as part shown in Fig~\ref{fig:cfp}, on the left. This is the so called $k_T$-factorisation theorem:
\beq \label{offshell}
\sigma(x,Q^2) = \int \frac{dz}{z} \frac{d {\bf k}_T^2}{{\bf k}_T^2}
C\left( \frac{x}{z}, \frac{Q^2}{{\bf k}_T^2} \right){G}(z, {\bf k}_T^2)\,,
\eeq
where $C$ is interpreted as an off-shell coefficient function, while $ G$ is the $k_T$-dependent gluon Green's function. The resummation of small-$x$ logarithms is usually obtained by  taking $G$ as the solution of the BFKL equation. More specifically, one diagonalises the convolution in Eq.~(\ref{offshell}) by computing Mellin moments with respect to $x$ and $Q^2$, obtaining
\beq
\Sigma(N,M)= h(N,M) F(N,M)\,,
\eeq
where we have defined  
\beq h(N,M)=  M\int_0^1 dx x^{N-1} \int_0^\infty d { {\bf k}_T^2} { ({\bf k}_T^2)}^{M-1} C(x, \bf{k_T}^2)
\eeq 
and $F$ is the Mellin transformed of $G$ (divided by $M$). The evolution of $F$ then gives the pole condition $M=\gamma_s\left(\frac{\alpha_s}{N} \right)$, where $\gamma_s=\sum_kc_k \left(\frac{\as}{N}\right)^k$ is the BFKL anomalous dimension, which resums poles in $N$, i.e. logarithms of $x$. The resummed result in $N$ space is then given by:
\beq \label{Nspace_res}
\Sigma(N) = h\left(N, \gamma_s\left(\frac{\as}{N}\right) \right).
\eeq

In our approach we still start from Eq.~(\ref{offshell}) but rather than solving the BFKL equation for $G$, guided by the generalised ladder expansion for collinear factorisation~\cite{CFP}, we look at $G$ as the iteration of collinear safe kernels $\gamma$, as depicted in Fig.~\ref{fig:cfp}, on the right. 
We compute the cross-section for $n$ insertions of the kernel $\gamma$ and we subtract the first $n-1$ poles according to the $\msb$ prescription. We find:
\bea \label{msb_res}
\sigma_n\lp N, Q^2, \as \lp
\frac{\mu^2}{Q^2}\rp^\epsilon, \epsilon\rp =
\gamma\lp N, \as\lp\frac{\mu^2}{Q^2} \rp^\epsilon, \epsilon\rp
\int_0^\infty \frac{d \xi_n}{\xi_n^{1+\epsilon}} 
C\lp N, \xi_n, \as \lp\frac{\mu^2}{Q^2}\rp^\epsilon, \epsilon\rp
\times \nonumber\\
\frac{1}{(n-1)!}\frac{1}{\epsilon^{n-1}} 
\left[
\sum_i \frac{\gamma_i(N,\alpha_s,0) }{i}
\lp 1 - \lp \frac{\mu^2}{Q^2 \xi_n} \rp^{i\epsilon}
\frac{\gamma_i(N,\alpha_s,\epsilon)}{\gamma_i(N,\alpha_s,0)}\rp
\right]^{n-1},
\eea 
where $\xi=\frac{{\bf k}_T^2}{Q^2}$. The full result is the obtained summing over $n$
\beq \label{our_res}
\sigma(N, Q^2, \as) =  \sum_n \sigma_n =\gamma(N,\as)
\int_0^\infty d\xi \xi^{\gamma(N,\as)-1} 
C(N,\xi, Q^2, \as)  {R}(N,\alpha_s).
\eeq
This is the same as the one in Eq.~(\ref{Nspace_res}), the only difference being the scheme dependent factor $R$, which reflects the fact that the calculation has been performed in $\msb$. The non-trivial information is encoded in the kernel $\gamma$. This kernel is an anomalous dimension, in the sense that it is the residue of a  collinear pole.  High-energy resummation is then achieved by choosing $\gamma$ to be the dual~\cite{duality} of the BFKL kernel. At the first non trivial order then $\gamma=\gamma_s$.

The main ingredient for obtaining Eq.~(\ref{msb_res}) is the use of the same kinematics as in the proof of collinear
factorisation~\cite{CFP}. Within this kinematics, the
dependence on transverse and longitudinal momentum components are 
kept separate from each other, and this rendered the generalisation to
rapidity distributions possible. 

A simple expression for the all-order rapidity distribution in the high energy limit can be obtained for the Fourier-Mellin transformed of the partonic rapidity distribution:
\beq
\frac{\d\sigma}{\d y}(N,b)
\equiv \int d x~ x^{N-1} \int d y~ e^{i b y} \frac{\d\sigma}{\d y}(x,y)\,.
\eeq
The resummed result is then a simple generalisation of the inclusive case
\bea \label{yres}
\frac{\d\sigma}{\d y}(N,b) &=&
\int_0^\infty d\xi_1
\gamma_s\lp N+ \frac{i b}{2} \rp
\xi_1^{{\gamma_s\lp N+ \frac{i b}{2} \rp}-1} 
\times \nonumber \\
&\times&
\int_0^\infty d\xi_2
\gamma_s\lp N- \frac{i b}{2} \rp
\xi_2^{\gamma_s\lp N- \frac{i b}{2} \rp-1} 
C(N,\xi_1,\xi_2, b).
\eea
We note that the argument of the anomalous dimension $\gamma_s$ is shifted by $\pm i b/2$. The off-shell coefficient function $C$ is now differential in the partonic rapidity.

\section{Higgs rapidity distribution}
We have applied the formalism described in the previous section to Higgs production in gluon-gluon fusion. High-energy factorisation can be used to compute the small-$x$ behaviour of the coefficient function keeping the full top mass dependence. By matching this result to an asymptotic expansion at large $x$, finite top mass effects have been evaluated to NNLO for the inclusive cross-section~\cite{SimoneHarlander, harlander, pak}. In the same spirit we construct an approximate NLO rapidity distribution by matching the result determined from high-energy factorisation with full top mass dependence at small-$x$ to the one obtained in the infinite top mass limit at large $x$~\cite{AnastasiouHiggsNLO}. 
Before doing that we can use the analytic results for the NLO rapidity distribution in the heavy top limit to check our method. We apply our main result Eq.~(\ref{yres}) to the case of Higgs production;  expanding it to NLO and inverting the Fourier-Mellin transform, we find
\beq
\frac{\d \sigma}{\d u} (x,u) = 
3\sigma_0 \frac \as \pi
\left[\frac{1}{(u-x)_+}-\delta(u-x)\ln x + \lp u \leftrightarrow \frac 1 u 
\rp \right]\quad {\rm with}\quad u=e^{-2 y},
\eeq
which is in full agreement with Ref.~\cite{AnastasiouHiggsNLO}. Similarly we can compute the small-$x$ behaviour of the rapidity distribution keeping the top mass finite:
\beq
\frac{\d \sigma}{\d u} = \sigma_0(\tau) c_1(\tau) \delta(u-x) + 
\lp u \leftrightarrow \frac 1 u \rp\quad {\rm with}\quad \tau \equiv \frac{4 m_t^2}{m_h^2},
\eeq
where $c_1$ has been determined numerically~\cite{CFM}.

\begin{figure}
\begin{center}
\includegraphics[height=.25\textheight]{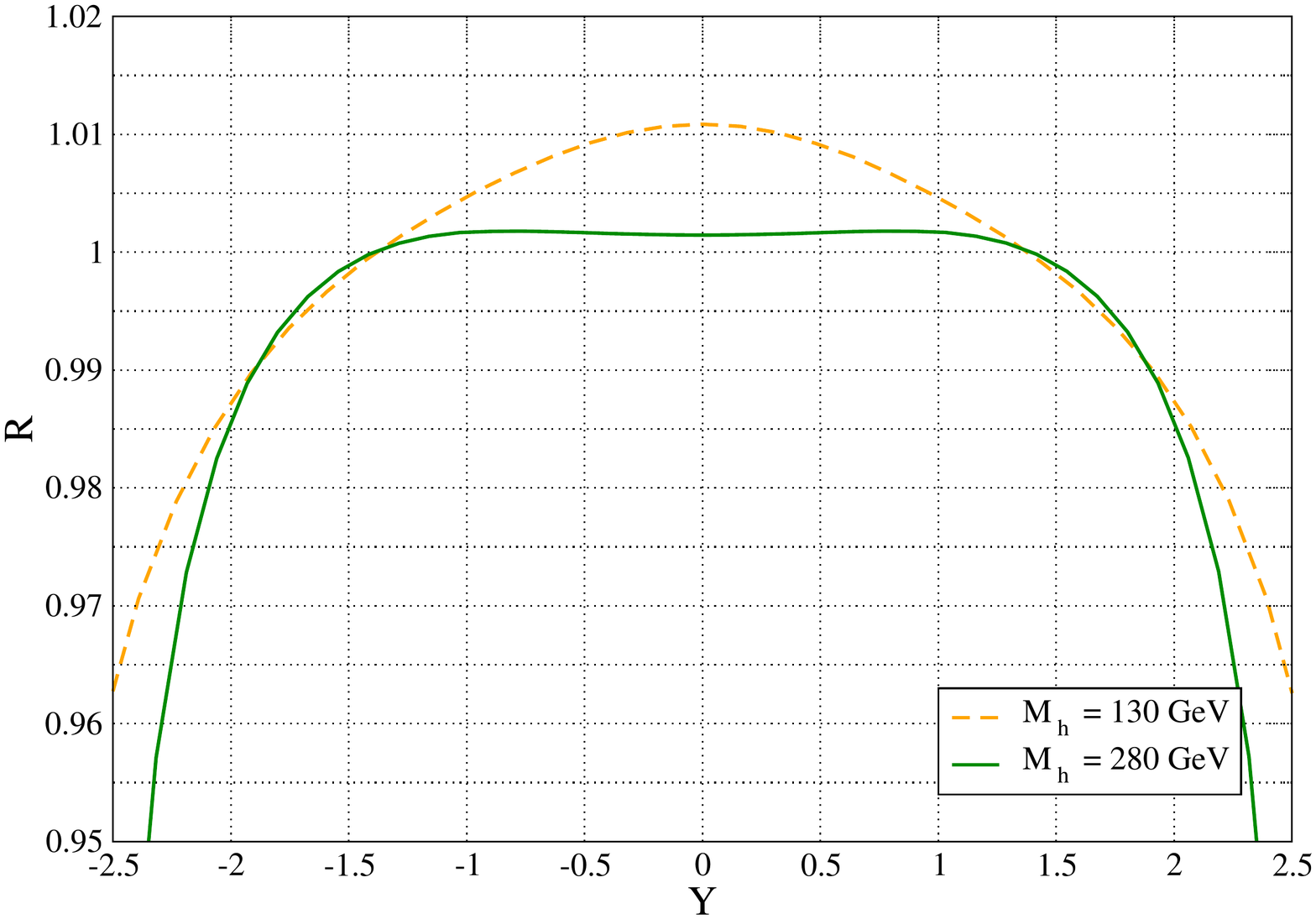}
\includegraphics[height=.25\textheight]{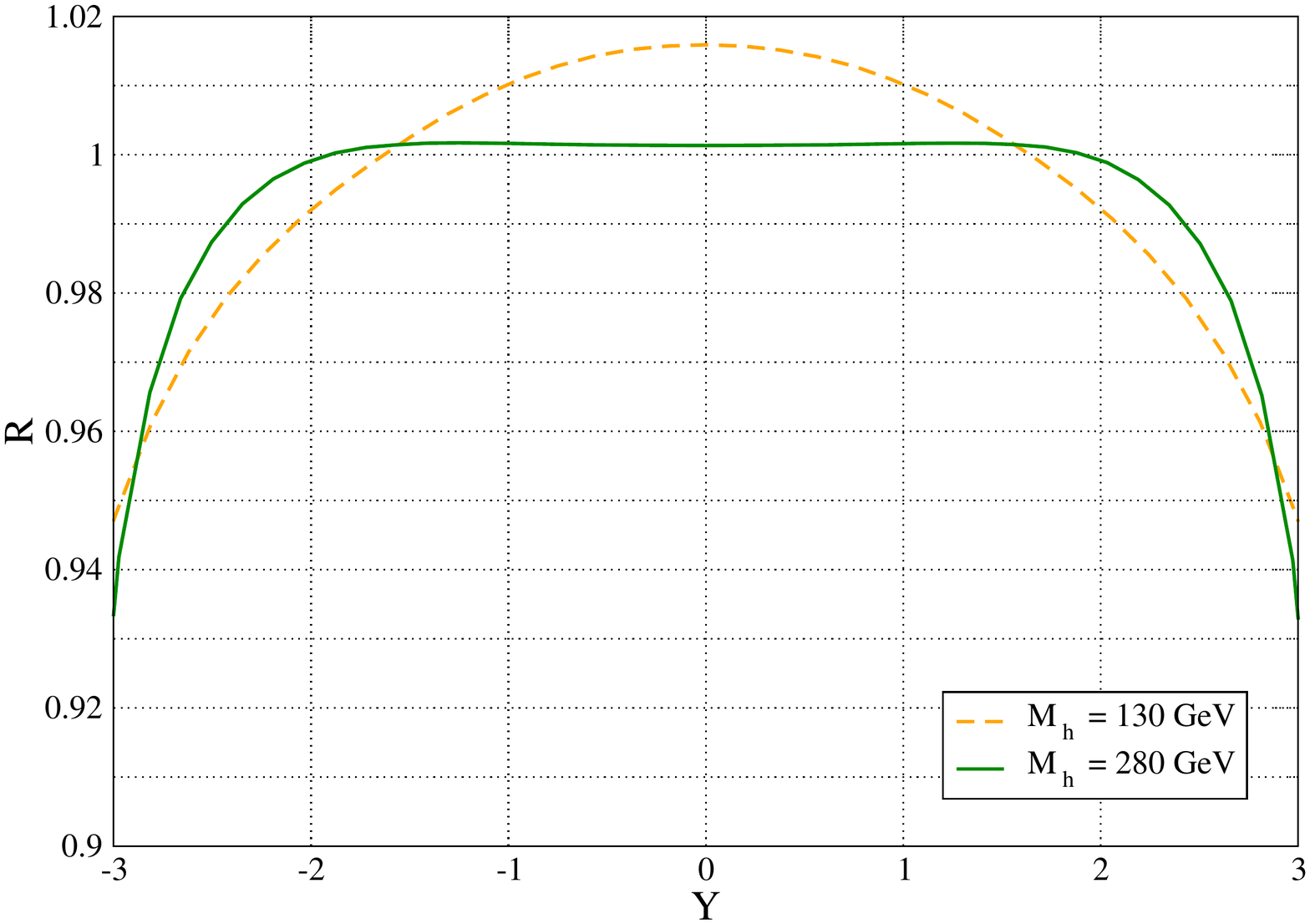}
\caption{The ratio $R$ as defined in Eq.~(\ref{Kfact}) for two different values of the Higgs mass $m_H=130$, $280$~GeV. The plot on the left is for $\sqrt{S}=7$~TeV, while the one on the right for $\sqrt{S}=14$~TeV}\label{fig:higgs}
\end{center}
\end{figure}

We study finite top mass effects by computing
\beq \label{Kfact}
R= \lp \frac 1 \sigma_{NLO} \frac{\d \sigma_{NLO}}{\d Y} \rp_{\rm matched}
\bigg / 
\lp \frac 1 \sigma_{NLO} \frac{\d \sigma_{NLO}}{\d Y} \rp_{m_t\rightarrow \infty}.
\eeq
Our findings are plotted in Fig.~\ref{fig:higgs} for proton-proton collisions at $\sqrt{S}=7$~TeV on the left and $\sqrt{S}=14$~TeV on the right for two different values of the Higgs mass. These plots confirm
the conclusion of Ref.~\cite{hpro} that corrections to the NLO
rapidity
distribution due to finite top mass effects are below 5\%.

\section{Conclusions and Outlook}
We have reported on an extension of high-energy factorisation to rapidity distributions. 
The result  has been made possible by exploiting the duality
which relates DGLAP and BFKL evolution equations, to re-express
high energy factorisation in terms of standard collinear
factorisation. As a first application, we have performed the resummation of the rapidity
distribution for Higgs production in gluon-gluon fusion.
More interesting application of this formalism is to
Drell-Yan rapidity distributions, which will be explored at very small
$x$ at the LHC.


\begin{thebibliography}{99}


\bibitem{CataniHQ}
  S.~Catani, M.~Ciafaloni and F.~Hautmann,
  Nucl.\ Phys.\  B {\bf 366} (1991) 135.


\bibitem{CataniDIS}
  S.~Catani and F.~Hautmann,
  Nucl.\ Phys.\  B {\bf 427} (1994) 475
  [arXiv:hep-ph/9405388].


\bibitem{BallEllis}
  R.~D.~Ball and R.~K.~Ellis,
  JHEP {\bf 0105} (2001) 053
  [arXiv:hep-ph/0101199].


\bibitem{HautmannHiggs}
  F.~Hautmann,
  Phys.\ Lett.\  B {\bf 535}, 159 (2002)
  [arXiv:hep-ph/0203140].


\bibitem{SimoneHiggs}
  S.~Marzani, R.~D.~Ball, V.~Del Duca, S.~Forte and A.~Vicini,
  Nucl.\ Phys.\  B {\bf 800}, 127 (2008)
  [arXiv:0801.2544 [hep-ph]].


\bibitem{SimoneHiggsProc}
  S.~Marzani, R.~D.~Ball, V.~Del Duca, S.~Forte and A.~Vicini,
  Nucl.\ Phys.\ Proc.\ Suppl.\  {\bf 186} (2009) 98
  [arXiv:0809.4934 [hep-ph]].


\bibitem{pseudo}
  F.~Caola and S.~Marzani,
  Phys.\ Lett.\  B {\bf 698} (2011) 275
  [arXiv:1101.3975 [hep-ph]].

\bibitem{SimoneDY}
  S.~Marzani and R.~D.~Ball,
  Nucl.\ Phys.\  B {\bf 814}, 246 (2009)
  [arXiv:0812.3602 [hep-ph]].


\bibitem{GiovanniPhoton}
  G.~Diana,
  Nucl.\ Phys.\  B {\bf 824}, 154 (2010)
  [arXiv:0906.4159 [hep-ph]].


\bibitem{CFM}
  F.~Caola, S.~Forte and S.~Marzani,
  Nucl.\ Phys.\  B {\bf 846} (2011) 167
  [arXiv:1010.2743 [hep-ph]].

\bibitem{duality}
 G.~Altarelli, R.~D.~Ball and S.~Forte,
  Nucl.\ Phys.\  B {\bf 575}, 313 (2000).


\bibitem{CFP}
  G.~Curci, W.~Furmanski and R.~Petronzio,
  Nucl.\ Phys.\  B {\bf 175} (1980) 27.



\bibitem{SimoneHarlander}
  R.~V.~Harlander, H.~Mantler, S.~Marzani and K.~J.~Ozeren,
  arXiv:0912.2104 [hep-ph].


\bibitem{harlander}
  R.~V.~Harlander and K.~J.~Ozeren,
  JHEP {\bf 0911} (2009) 088
  [arXiv:0909.3420 [hep-ph]].


\bibitem{pak}
  A.~Pak, M.~Rogal and M.~Steinhauser,
  JHEP {\bf 1002} (2010) 025
  [arXiv:0911.4662 [hep-ph]].

\bibitem{AnastasiouHiggsNLO}
  C.~Anastasiou, L.~J.~Dixon and K.~Melnikov,
  Nucl.\ Phys.\ Proc.\ Suppl.\  {\bf 116} (2003) 193
  [arXiv:hep-ph/0211141].

\bibitem{hpro}
  C.~Anastasiou, S.~Bucherer and Z.~Kunszt,
  JHEP {\bf 0910} (2009) 068
  [arXiv:0907.2362 [hep-ph]].


\end{thebibliography}
\end{document}